\documentclass[aps,twocolumn,amsmath,amssymb,groupedaddress,longbibliography]{revtex4-2}
\usepackage{bbm}
\usepackage{mathrsfs}
\usepackage{amsmath}
\usepackage{amsfonts}
\usepackage[colorlinks=true,citecolor=blue,anchorcolor=blue]{hyperref}
\usepackage{graphicx,epstopdf}
\usepackage{subfigure}
\usepackage{epsfig}
\usepackage{dcolumn}
\usepackage{bm}
\usepackage{color}
\usepackage{natbib}
\usepackage{amssymb}
\usepackage{xcolor}
\usepackage{braket}
\usepackage{soul}

\begin{document}

\title{Interaction-induced multiparticle bound states in the continuum}

\author{Boning Huang $^{1,2,3}$}

\author{Yongguan Ke $^{1,2,3,4}$} \email{Corresponding author. Email: keyg3@mail.sysu.edu.cn}

\author{Honghua~Zhong $^{5}$}

\author{Yuri S. Kivshar$^{6}$}

\author{Chaohong Lee $^{1,2,3,7}$} \email{Corresponding author. Email: chleecn@szu.edu.cn}

\affiliation{$^{1}$Institute of Quantum Precision Measurement, State Key Laboratory of Radio Frequency Heterogeneous Integration, Shenzhen University, Shenzhen 518060, China}

\affiliation{$^{2}$College of Physics and Optoelectronic Engineering, Shenzhen University, Shenzhen 518060, China}

\affiliation{$^{3}$Laboratory of Quantum Engineering and Quantum Metrology, School of Physics and Astronomy, Sun Yat-Sen University (Zhuhai Campus), Zhuhai 519082, China}

\affiliation{$^{4}$International Quantum Academy, Shenzhen 518048, China}

\affiliation{$^{5}$Institute of Mathematics and Physics, Central South University of Forestry and Technology, Changsha, 410004, China}

\affiliation{$^{6}$Nonlinear Physics Center, Australian National University, Canberra ACT 2601, Australia}

\affiliation{$^{7}$Quantum Science Center of Guangdong-Hongkong-Macao Greater Bay Area (Guangdong), Shenzhen 518045, China}

\date{\today}
	
\begin{abstract}
Bound states in the continuum (BICs) are localized modes residing in the radiation continuum. They were first predicted for single-particle states, and became a general feature of many wave systems. In many-body quantum physics, it is still unclear what would be a close analog of BICs, and whether interparticle interaction may induce BICs. Here, we predict a novel type of multiparticle states in the interaction-modulated Bose-Hubbard model that can be associated with the BIC concept. Under periodic boundary conditions, a so-called quasi-BIC appears as a bounded pair residing in a standing wave formed by the third particle. Under open boundary conditions, such a hybrid state becomes an eigenstate of the system. We demonstrate that the Thouless pumping of the quasi-BICs can be realized by modulating the on-site interactions in space and time. Surprisingly, while the center-of-mass of the quasi-BIC is shifted by a unit cell in one cycle, the bounded pair moves into the opposite direction with the standing wave.
\end{abstract}


\maketitle
%

%
Bound states in continuum (BICs) are known as spatially localized states residing in the continuum spectrum of radiative states~\cite{neumann1929merkwurdige,hsu2016bound,PhysRevB.100.115303,Koshelev:20,azzam2021photonic}.
In practice, an ideal BIC being a dark mode with infinite lifetime always turns into a quasi-BIC with finite lifetime~\cite{sadrieva2017transition,doeleman2018experimental,PhysRevLett.121.193903}.
The study of BIC and quasi-BIC in photonic systems has attracted intense interest because of the fundamental mechanism they provide for many problems of light-matter interaction~\cite{PhysRevB.98.161113,cao2020normal,kravtsov2020nonlinear} and important applications such as lasers~\cite{meier1999laser,imada1999coherent,kodigala2017lasing,rybin2017supercavity,huang2020ultrafast}, high-harmonic generation~\cite{PhysRevLett.121.033903,koshelev2019nonlinear,koshelev2020subwavelength,Zograf:20}, sensing and imaging~\cite{liu2017optical,leitis2019angle,yesilkoy2019ultrasensitive}.
The mechanics of generating BIC vary from symmetry protection~\cite{PhysRevLett.102.167404,PhysRevLett.107.183901}, separable subspace~\cite{rivera2016controlling}, destructive interference~\cite{PhysRevB.78.075105,gao2016formation}, Fabry-P\'erot cavity~\cite{PhysRevLett.100.183902,PhysRevB.78.075105}, time-periodic modulation~\cite{gonzalez2013bound,longhi2013floquet,PhysRevB.96.104309,PhysRevA.102.023303}, to inverse construction~\cite{PhysRevLett.108.070401,gallo2014bulk}.
Single-particle BICs can be protected by a topological charge being robust to a variation of the system parameters~\cite{yang2013topological,PhysRevLett.113.257401}.
Tunability and dynamic control of BICs can be achieved via optical pumping~\cite{fan2019dynamic,han2019all}, the use of phase-change materials~\cite{mikheeva2019photosensitive,chen2020tunable}, which are of interest due to potential applications.

Beyond conventional BICs for classical waves and single-particle quantum systems, in the presence of impurities, multiparticle quantum BICs may exist in one-dimensional Hubbard systems~\cite{PhysRevLett.109.116405,Zhang2013,Hong-HuaZhong:70304,sugimoto2023manybody} and may prevent the system from thermalization~\cite{sugimoto2023manybody}.
Even without impurity, two-particle BICs appear in a Hubbard lattice with anyonic statistics~\cite{zhang2023anyonic} or  an intense oscillating electric field~\cite{PhysRevB.89.115118}.
Up to now, it is still unclear whether particle-particle interactions may induce BICs and there are no studies of dynamical control of multiparticle BICs.
On the other hand, although it has been theoretically predicted and experimentally demonstrated the topological Thouless pumping of conventional bound states~\cite{PhysRevA.95.063630,walter2023quantization,tao2023interactioninduced}, it is still challenging and appealing to achieve the topological Thouless pumping of multiparticle BICs.

\begin{figure}[t]
    \centering
    \includegraphics[scale=0.26]{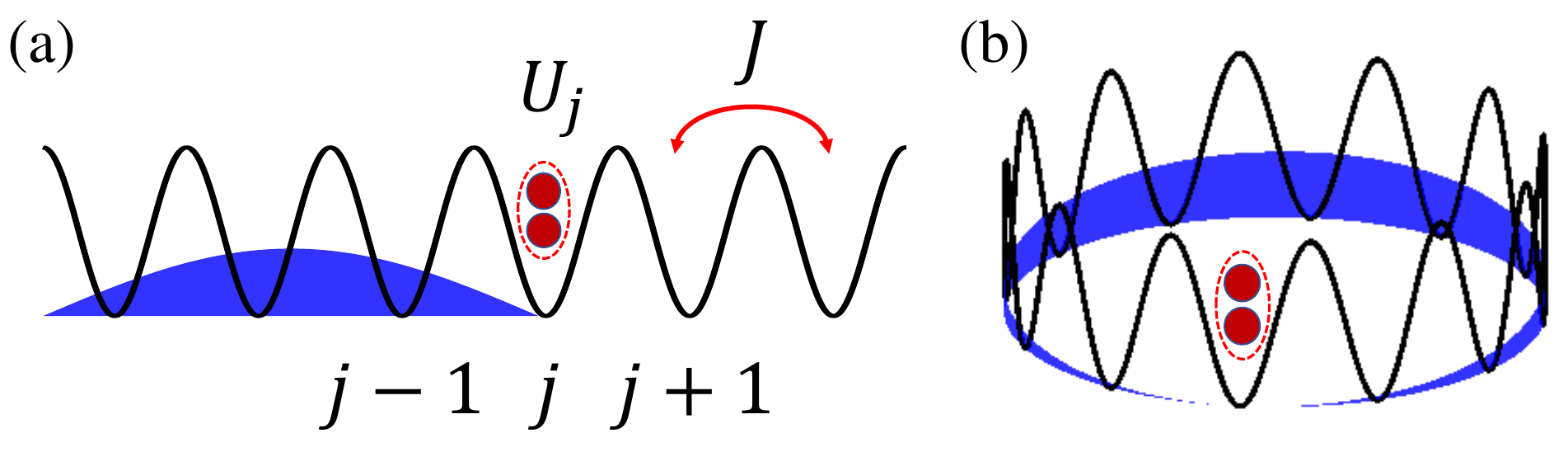}
    \caption{Schematics for (a) bound state in continuum under open boundary condition and (b) quasi-bound state in continuum under periodic boundary condition.}
    \label{Schematic}
\end{figure}

In this Letter, we predict the existence of few-particle BICs and quasi-BICs in a Bose-Hubbard model without impurity and propose how to achieve topological pumping of quasi-BICs by modulating particle-particle interactions in space and time.
Under open boundary condition, multiparticle BIC of three particles appears as a bound pair localized by the standing wave of the other particle in a finite lattice; see Fig.~\ref{Schematic}(a). 
Under periodic boundary condition, we develop a new method to construct multiparticle maximally localized Wannier states (MLWS) via the projected position operator. 
We find that quasi-BIC, constructed from three-particle MLWS, appears as a localized bound pair in the standing wave of the other particle; see Fig.~\ref{Schematic}(b). 
In the pumping process, the transport direction of the standing wave is opposite to the one of the bounded pair.
While the topological pumping of BIC breaks down, the mean position of quasi-BIC is shifted anticlockwise by one unit cell.
The displacement is directly related to the Chern number of the multiparticle band which is uniformly populated by the quasi-BIC.
Our work paves the way for systematic generation and topological control of the multiparticle BICs.


Depicted in Fig.~\ref{Schematic}(a) under open boundary condition and Fig.~\ref{Schematic}(b) under periodic boundary condition, we consider interacting bosons in a lattice with spatial modulation of onsite interaction,
%

\begin{equation}
        \hat{H}=J\sum_{j}(\hat{a}_{j+1}^{\dag}\hat{a}_j+h.c.)+\frac{1}{2}\sum_{j}U_j(\phi)\hat{n}_j(\hat{n}_j-1).
\end{equation}
Here, $\hat{a}_{j}^{\dag}$ ($\hat{a}_j$) and $\hat{n}_j=\hat{a}_{j}^{\dag}\hat{a}_j$ are the operators of bosonic creation (annihilation) and particle number at the $j$th site, respectively.
The system size is $M$.
$J$ is the nearest neighboring hopping strength. 
$U_j=U_0+\delta \cos(2\pi\beta j+\phi)$ is the  on-site interaction strength, where $\delta$, $\beta=p/q$ ($p$, $q$ being co-prime numbers),  and $\phi$ are the modulation strength, spatial frequency and modulation phase, respectively. 
$U_j$ can be tuned by varying the modulation phase as $\phi=\omega t$, where $\omega$ is the temporally modulation frequency with modulation period $T=2\pi/\omega$. 
In this work, we concentrate on the strong interaction regime ($|U|>>|J|$), and set $\beta=1/3$ without loss of generality. 
Such a model could be readily realized in an array of superconducting circuits~\cite{tao2023interactioninduced}.
The onsite interaction can be independently and dynamically tuned by controlling the anharmonicity of individual qubits.
If the two particles are separately located at different sites, they can tunnel independently without onsite energy.
If two more particles are at the same site, due to the strong onsite interaction, they form bound states which tunnel as a whole~\cite{winkler2006repulsively} and feel a periodic onsite energy with period $q=3$, described by an effective Aubry-Andr\'e-Harper model~\cite{Supplementary}.
Thus, the spatiotemperal modulation only breaks time reversal symmetry of bound states but plays no role for a single particle.
\begin{figure*}
    \centering
\includegraphics[scale=0.42]{BIC_M30.pdf}
    \caption{BIC under open boundary condition and quasi-BIC under periodic boundary condition.
    (a) Energy spectrum at $\phi=0$ under open boundary condition, and the inset shows the bound state in continuum given by the type-(ii) eigenstates with energy $\epsilon=321.9826$(left) and $\epsilon=320.8269$(right), which are indicated by the orange circle and cross in the spectrum respectively.
    (b) Multiparticle Bloch bands at $\phi=0$, and the inset shows the quasi-bound state in continuum given by the maximally localized Wannier state of the highest (left) and middle (right) type-(ii) bands at $t=0$, which are indicated by the orange asterisk and plus sign in the bands respectively.
    (c) Decay ratio $D=\langle\hat{n}_j\rangle(t)/\langle\hat{n}_j\rangle(0)$ as a function of time, where the bounded two bosons locating at the $j$-th site. The green, blue solid lines, green dashed line and blue dotted line correspond to the four (quasi-)BICs in the insets of Fig.\ref{BIC}(a) and (b), respectively. 
    The parameters are $N=3, M=30, U_0=300, \delta=20, J=1$.
    }
    \label{BIC}
\end{figure*}


First, we first consider three-boson eigenvalues and eigenstates at $\phi=0$ under the open boundary condition and periodic boundary condition; see the left and middle panels of Fig.~\ref{BIC}, respectively.
Under the periodic boundary condition the system has co-translational symmetry, that is, the energy is conserved when all particles are shifted as a whole by unit cells. Therefore, the center of mass momentum $\kappa$ is a good quantum number, and the energy spectrum as a function of $\kappa$ forms multiparticle energy bands.
Due to the strong repulsive interaction ($U/J=300$), the energy spectrum [Fig.~\ref{BIC} (a)] and the multiparticle energy band [Fig.~\ref{BIC}(b)] are separated into three clusters. 
The other parameters are chosen as $J=1$, $\delta=20$, and $M=30$.
From the bottom to the top, the three clusters (marked by red, blue, and green) correspond to (i) three independent bosons, (ii) two bound bosons and an independent bosons, and (iii) three bound bosons.
Since the type-(i,iii) states are well known~\cite{Supplementary}, we focus on the type-(ii) states.
We notice that in the conventional Bose-Hubbard model both the bound pair and independent bosons in the type-(ii) states are always extended over the lattice.
Surprisingly, in our systems under open boundary we find all type-(ii) states are a localized bound pair by a standing wave; see insets of Fig.~\ref{BIC}(a) for two typical eigenstates with energies $\epsilon=321.9826$ (left) and $\epsilon=320.8269$ (right) and their higher-order correlation functions in the Supplemental Materials~\cite{Supplementary}.
Considering the cluster of type-(ii) bound states forms the continuum spectrum as the system size increases, we term the localized type-(ii) states under the open boundary condition as multiparticle bound state in continuum.
Since three-paticle BICs can be decomposed to an extended single-particle state and a localized two-particle state, we can obtain an effective Hamiltonian for the localized part of some BICs~\cite{Supplementary}  
\begin{equation}
\begin{aligned}
    \hat{H}_{\rm eff}= \hat H+\sum_{j}2U_j(\phi){|\varphi_j|}^2\hat{n}_j,
\end{aligned}
\end{equation}
where $\varphi_j$ is the amplitude of the extended single-particle state, which provides background potential.
The idea of background potential provided by the particle itself can explain a variety of localization phenomena ~\cite{kagan1984localization,Grover_2014,PhysRevB.90.165137,PhysRevLett.117.240601,PhysRevLett.124.093604,PhysRevA.103.023720,PhysRevB.104.195102,kirsch2021nonlinear,PhysRevB.107.045131,Zhang_2023}.

There is no multiparticle BIC under periodic boundary condition because the systems are translation invariant. 
However, we find that the multiparticle energy bands for the type-(ii) states are flat because of the strong interaction.
The flat energy band means that the particles are almost localized as the group velocities of the multiparticle Bloch states $|\psi_m(\kappa)\rangle$ ($m\in $ the type-(ii) states) almost vanish.
We can construct multiparticle maximally localized Wannier states (MLWS) of the flat band which are approximately the eigenstates.
To this end, we first make use of the projected position operator $\hat{P}\hat{x}\hat{P}$, 
\begin{equation}
\hat{P}=\sum_{m\in\mathcal{M},\kappa}|\psi_m(\kappa)\rangle\langle \psi_m(\kappa)|,
\end{equation}
where position operator is given by $\hat x=N^{-1}\sum_j j \hat{n}_j$, and $m$ takes targeted $\mathcal M$ multiparticle Bloch bands.
Similarly to single-particle MLWS~\cite{PhysRevB.56.12847}, we choose the multiparticle Wannier state $|W_m(0)\rangle$ as the eigenstate of the projected position operator with eigenvalue $X_{0m}$,
that is,
\begin{equation}
\begin{aligned}
     \langle W_n(R)|\hat{x}|W_m(0)\rangle&=\langle W_n(R)|\hat{P}\hat{x}\hat{P}|W_m(0)\rangle
     \\&=X_{0m}\delta_{R,0}\delta_{m,n},
\end{aligned}
\end{equation}
where $m,n\in \mathcal{M}$. We can prove that $|W_m(0)\rangle$ is the multiparticle MLWS~\cite{Supplementary}.
As the system size increases, the energy gaps between these states vanish and the spectrum becomes continuum.
We term the multiparticle MLWS of the type-(ii) band as multiparticle quasi-bound state in continuum (quasi-BIC). 
Insets of Fig.~\ref{BIC}(b) show the density distribution $\langle n_j\rangle$ of two quasi-BICs in the highest (left) and middle (right) of the (ii) type cluster bands; see higher order correlation functions in the Supplemental Materials~\cite{Supplementary}. 
These multiparticle quasi-BICs are a bound pair in the standing wave of the other particle.

Theoretically, BICs are eigenstates which have infinite lifetimes, and quasi-BICs are approximately eigenstates which have long lifetimes.
By setting the initial states as the BIC and quasi-BIC in insets of Figs.~\ref{BIC}(a) and (b), the difference between BIC and quasi-BIC can be found by tracing the ratio $D=n_j(t)/n_j(0)$ in a long time evolution under the static Hamiltonian with $\phi=0$, where the bound pair is located at the $j$-th site; see Fig.~\ref{BIC}(c).
We find $D=1$ preserves in all time for BIC and $D\approx 1$  even at the time scale of $Jt\approx 10^4$ for the right quasi-BIC in Fig.~\ref{BIC}(b).
The left quasi-BIC has a much smaller decay rate than the right quasi-BIC, because it comes from the energy band which is more flat, and the BICs have no decay.

\begin{figure*}
    \centering
    \includegraphics[width=0.9\textwidth]{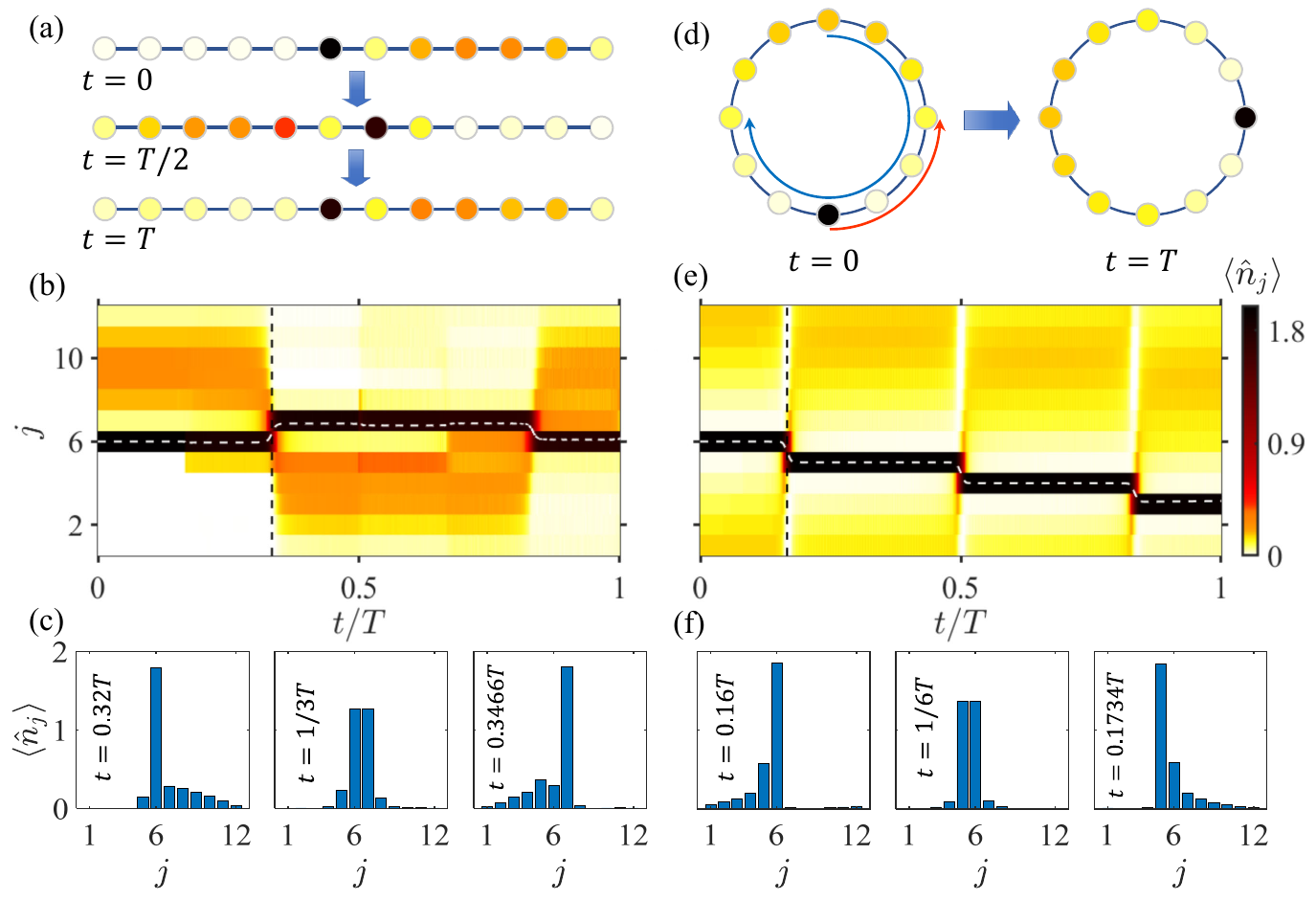}
    \caption{Pumping dynamics of BIC (left) and quasi-BIC (right). (a,d) Schematics for the dynamical localization of BIC and topological pumping of quasi-BIC. Colors denote the density distribution. 
    (b,e) Density distribution as a function of time for (a,d). The white dashed line denotes the c.m. position of the bound pair. 
    (c,f) Density distribution of three instantaneous states around transition time $t=1/3T$ for BIC and $t=1/6T$ and quasi-BIC. 
    (d) Schematics for the adiabatic evolution of BIC under open boundary condition in one cycle in a ring lattice with $12$ sites. Colors denote the density distribution of the initial quasi-BIC (left) and the final quasi-BIC (right). 
    (e) Density distribution as a function of time for (d). The white dashed line denotes the c.m. position of the bound pair.
    (f) Density distribution of two instantaneous states around the transition moment in (e). 
    The parameters are set as $U_0=300, \delta=20, J=1, \omega=0.0005$. }
    \label{Pumping}
\end{figure*}

%
Based on multiparticle Thouless pumping theory~\cite{PhysRevA.95.063630}, if an initial state $|\psi(0)\rangle$ uniformly fills and sweeps a multiparticle energy band with nontrivial Chern number, 
\begin{equation}
C_m=\frac{i}{2\pi} \int d\kappa \int  dt (\langle\partial_t\psi_m|\partial_\kappa \psi_m\rangle-\langle\partial_\kappa\psi_m|\partial_t \psi_m\rangle),
\end{equation}
the position shift of the multiparticle Wannier state in one cycle ($T=2\pi/\omega$) is equal to the Chern number multiples of unit cell,
$\Delta x(T) = \langle \psi(T)|\hat x|\psi(T) \rangle- \langle \psi(0)|\hat x|\psi(0) \rangle=  C_m q$.
The Chern numbers of the highest and lowest type-(ii) band are $-1$, and there are finite energy gaps that separate other energy bands, which make it possible to perform topological pumping.


We consider the particle transport for BIC and quasi-BIC under a spatiotemperal modulation of interaction. 
Under open boundary condition, the initial state is set as the multiparticle BIC given by the type-(ii) eigenstate with energy $\epsilon=321.8212$, and it evolves with time as $|\psi(t)\rangle=\mathcal{T}\exp [-i\int_0^t \hat H(\tau)d\tau]|\psi(0)\rangle$, where $\mathcal {T}$ is time ordering operator.
The numerical studies of the dynamics require a lot of computational resources for large $M$ and $N$ and discrete time sequences.
However, strong on-site interaction supports large energy gaps between the types (ii) and (i,iii) states. 
%
%
Because of negligible coupling between states of different types under slow modulation, we can only consider the Hamiltonian in the subspace $\mathcal V=\{|0,....,n_i=2,...,n_j=1,...\rangle \}$, that reduces the dimension of the Hilbert space from $(M+2)(M+1)M/6$ to $M(M-1)$.
The method of subspace projection is beneficial to numerical simulations of many-body systems~\cite{Supplementary}. 
By choosing parameters as $U_0=300, \delta=20, J=1, \omega=0.0005$ and $M=12$, we calculate the density distribution $\langle \hat n_j(t)\rangle=\langle\psi(t)|\hat n_j|\psi(t)\rangle$; see Fig.~\ref{Pumping}(a) and ~\ref{Pumping}(b) for the schematic and time evolution, respectively.
At the $6$th site, the bound pair tunnels to the $7$th site and returns to the $6$th site in one cycle; see a dashed white line for the mean position of the bound pair. 
Transitions occur mainly at $t=1/3T$ and $t=5/6T$.
Clearly, BIC is dynamically localized at multiples of the pumping cycle, and the topological pumping of the multiparticle BIC breaks down because there is no uniform band occupation.
To show the motion of the other particle,  we calculate the density distribution before, at, and after the transition time $t=1/3T$; see left, middle, and right panels in Fig.~\ref{Pumping}(c), respectively.
We find that the standing wave gradually gathers around the right-hand side of the bound pair before the transition, it mixes with the bound pair at the transition, and then it tunnels to the left-hand side, after the transition. %
The transport direction of the standing wave is opposite to that of the bound pair.

However, employing periodic boundary condition, we can realize topological pumping of a quasi-BIC filling uniformly the highest band of the type (ii), see Fig.~\ref{Pumping}(d) for schematic and \ref{Pumping}(e) for the time evolution of the density distribution.
The initial state is composed of a two-particle bound state localized at the $6$th site and a standing wave of the other particle centered at the $12$th site.
Under slow modulation, the initial state sweeps the highest band of the type (ii), and the global Berry curvatures play an important role.
Unidirectional tunnelings of the bound pair occur mainly at $t=T/6,~T/2,~5T/6$.
We find that in one cycle the mean position of the bound pair is shifted from the $6$th to the $3$th sites; see a white dashed line in Fig.~\ref{Pumping}(e).
The left and right panels of Fig.~\ref{Pumping}(d) show almost the same profiles of the initial and final states,
except that the relative distance between the bound pair and the standing wave is shifted by $0.0103$ unit cell. 
The mean position is anticlockwise being shifted by $-0.9896$ unit cell, that is consistent with the Chern number $C=-1$ of the filled band.
To understand the corresponding microscopic picture, we show the density distribution around the transition time at $t=1/6T$; see Fig.~\ref{Pumping}(f). 
We find that the standing wave first gathers around the left side of the bound pair and then mixes with the two particles at $t=T/6$, and crosses to the right side of the bound pair.
The mixed state at $t=T/6$ is presented approximately by $(|1_5,2_6\rangle+|2_5,1_6\rangle)/\sqrt{2}$, which can be prepared by local excitation.
This means that the standing wave moves clockwise in the pumping process, which is opposite to the anticlockwise direction of the bound-pair transport. 
This mechanics differs substantially the topological pumping of bound states and topological resonant tunnelings, where all particles move in the same direction as a whole or one by one~\cite{PhysRevA.95.063630,walter2023quantization,tao2023interactioninduced}.
We notice that although the energy gaps between different quasi-BICs are small, the topological pumping of quasi-BICs is relatively robust to disorder in the onsite energies~\cite{Supplementary}.

In summary, we have suggested a systematic method for generating BIC and quasi-BICs, that can be extended to a variety of many-body systems, involving fermionic particles, spin chain, and hybrid quantum systems.
We notice that, as the next step, it is interesting to construct maximally localized multiparticle Wannier states in two dimensions, which can support richer many-body quasi-BICs.
While the Thouless pumping provides a simple way to control the spatial freedom of quasi-BICs in a topologically quantized way, some further studies are required to explore the possibility of fractional topological pumping of many-body quasi-BICs.
Many-body BICs have been proven to break ergodicity in many-body systems~\cite{sugimoto2023manybody}.
In our case, strong interaction is required to stabilize quasi-BICs. 
However, when the interaction strength decreases, a small group of quasi-BICs of the same band at different sites will couple with each other.
Our system may provide a new possibility to realize the Hilbert space fragmentation, many-body scar, and ergodicity breaking.

\begin{acknowledgments}
The authors acknowledge useful discussions with Alexander Poddubny, Wenjie Liu, Ling Lin, Li Zhang, and Na Zhang. This work was supported by the National Key Research and Development Program of China (Grant No. 2022YFA1404104), the National Natural Science Foundation of China (Grants Nos. 12025509, 12275365, and 12175315), the Key-Area Research and Development Program of Guangdong Province (Grant No. 2019B030330001), the Natural Science Foundation of Guangdong (Grant No. 2023A1515012099), and the Australian Research Council (Grant No. DP200101168).
\end{acknowledgments}

\nocite{*}

\providecommand{\noopsort}[1]{}\providecommand{\singleletter}[1]{#1}%

\onecolumngrid
\clearpage

\renewcommand {\Im}{\mathop\mathrm{Im}\nolimits}
\renewcommand {\Re}{\mathop\mathrm{Re}\nolimits}
\renewcommand {\i}{{\rm i}}
\renewcommand {\phi}{{\varphi}}

\begin{center}
	\noindent\textbf{\large{Supplemental Materials:}}
	\\\bigskip
	\noindent\textbf{\large{Topological inverse band theory in waveguide quantum electrodynamics}}
	\\\bigskip
	\onecolumngrid
	
	Boning Huang$^{1,2,3}$, Yongguan Ke$^{1,2,3,4,*}$, Honghua Zhong$^{5}$,  Yuri S. Kivshar$^{6}$, Chaohong Lee$^{1,2,3, 7 \dag}$
	
	\small{$^1$ \emph{Institute of Quantum Precision Measurement, State Key Laboratory of Radio Frequency Heterogeneous Integration, Shenzhen University, Shenzhen 518060, China}}\\
	\small{$^2$ \emph{College of Physics and Optoelectronic Engineering, Shenzhen University, Shenzhen 518060, China}}\\
	\small{$^3$ \emph{Laboratory of Quantum Engineering and Quantum Metrology, School of Physics and Astronomy, Sun Yat-Sen University (Zhuhai Campus), Zhuhai 519082, China}}\\
	\small{$^4$ \emph{International Quantum Academy, Shenzhen 518048, China}}\\
	\small{$^5$ \emph{Institute of Mathematics and Physics, Central South University of Forestry and Technology, Changsha, 410004, China}}\\
	\small{$^{6}$ \emph{Nonlinear Physics Center, Australian National University, Canberra ACT 2601, Australia}} \\
	\small{$^{7}$ \emph{Quantum Science Center of Guangdong-Hongkong-Macao Greater Bay Area (Guangdong), Shenzhen 518045, China}}
\end{center}

\setcounter{equation}{0}

\setcounter{figure}{0}

\setcounter{table}{0}
\renewcommand{\theequation}{S\arabic{equation}}

\renewcommand{\thefigure}{S\arabic{figure}}
\renewcommand{\theHfigure}{S\arabic{figure}}

\renewcommand{\thesection}{S\arabic{section}}


\section{Derivation of effective Hamiltonian for bound states}
In this section, we show how to derive effective Hamiltonian for three-particle bound states through perturbation theory~\cite{PhysRevA.90.062301S,M.Takahashi_1977S}.
When $U_0\gg(\delta,J)$, the term noted as 
\begin{equation}
	\begin{aligned}
		\hat{H}'=&J\sum_{j}(\hat{a}^{\dag}_{j+1}\hat{a}_j+h.c.)+\frac{1}{2}\sum_{j}\delta \cos(2\pi\beta j+\phi)n_j(n_j-1)
	\end{aligned}
\end{equation}
can be treated as a perturbation to the term noted as
\begin{equation}
	\hat{H}_0=\frac{1}{2}\sum_{j}U_0n_j(n_j-1).
\end{equation}
When there are three bosons, the Hilbert space is divided into the subspace $u$ expanded with degenerate eigenstates $|3\rangle_j$ of $\hat{H}_0$ with eigenvalues $E_j=3U_0$ and its complement subspace $\nu$ expanded by degenerate eigenstates $|2\rangle_j|1\rangle_k(j\neq k)$ with eigenvalues $E_{j,k}=U_0$ and degenerate eigenstates $|1\rangle_j|1\rangle_k|1\rangle_l(j\neq k \neq l)$ with eigenvalues $E_{j,k,l}=0$.
Here, $|3\rangle_j$, $|2\rangle_j|1\rangle_k$, and $|1\rangle_j|1\rangle_k|1\rangle_l$ are short for Fock states $|0,...,n_j=3,..., 0\rangle$, $|0,...,n_j=2,...,n_k=1,..., 0\rangle$, and $|0,...,n_j=1,...,n_k=1,...,n_l=1,...,0\rangle$, respectively.
The projection operators upon $u,\nu$ are respectively defined as 
\begin{equation} \label{Projection3}
	\begin{aligned}
		\hat{P}=&\sum_j|3\rangle_j\langle 3|_j,\\
		\hat{S}=\sum_{j\neq k}\frac{1}{E_j-E_{j,k}}|2\rangle_j|1\rangle_k\langle 1|_k\langle 2|_j&+\sum_{j\neq k,j\neq l, k\neq l}\frac{1}{E_j-E_{j,k,l}}|1\rangle_j|1\rangle_k|1\rangle_l\langle1|_l\langle1|_k\langle1|_j
	\end{aligned}
\end{equation}
Applying the degenerate perturbation theory up to the third order, the effective Hamiltonian of subspace $u$ is given by
\begin{equation} \label{HamEffGen3}
	\hat{H}_{\rm eff}=\hat{P}\hat{H}\hat{P}+\hat{P}\hat{H}'\hat{S}\hat{H}'\hat{P}+\hat{P}\hat{H}'\hat{S}\hat{H}'\hat{S}\hat{H}'\hat{P}-\hat{P}\hat{H}'\hat{S}^2\hat{H}'\hat{P}\hat{H}'\hat{P}.
\end{equation}
Substituting Eq.~\ref{Projection3} into Eq.~\ref{HamEffGen3}, the effective Hamiltonian is given by
\begin{equation}
	\hat{H}_{\rm eff}=\sum_{j}3\left[U_0+(1-\frac{J^2}{U_0^2})\delta \cos(2\pi\beta j + \phi)\right]\hat{c}_{j}^{\dag}\hat{c}_j+\frac{3J^3}{2U_0^2}\sum_{j}(\hat{c}_{j+1}^{\dag}\hat{c}_j+h.c.).
\end{equation}
Here, $\hat{c}_{j}^{\dag}=\frac{1}{\sqrt{6}}\hat{a}_{j}^{\dag}\hat{a}_{j}^{\dag}\hat{a}_{j}^{\dag}$ is the creation operator of three bosons as a whole at the $j$th site, and a uniform on-site energy shift $3J^2/U_0$ is neglected.
Similarly, the effective model can be extended to the case of $N$-particle bound state with the general form
\begin{equation}
	\hat{H}_{\rm eff}=\sum_{j}U_{\rm eff}\hat{d}_{j}^{\dag}\hat{d}_j+J_{\rm eff}\sum_{j}(\hat{d}_{j+1}^{\dag}\hat{d}_j+h.c.),
\end{equation}
with $\hat{d}_{j}^{\dag}=\frac{1}{\sqrt{N!}}(\hat{a}_j^{\dag})^N$. The effective onsite energy $U_{\rm eff}$ and hopping strength $J_{\rm eff}$ can be obtained in a similar way.
In the general case, the effective Hamiltonian is the well-known Aubry-Andr\'e-Harper model of a quasiparticle as a $N$-particle bound state.

\section{Dynamics of the type-(i,iii) states for three particles}
\begin{figure}[!h]
	\centering
	\includegraphics[scale=0.54]{BS_pumping_threeparticles.pdf}
	\caption{Density distribution as a function of time for a adiabatic evolution of (a) bound state and (b) scattering state and (c) evolution of same initial state under static Hamiltonian. The red dashed line in (a) denotes mean position of the bounded three bosons. The parameters are $M=12, U_0=30, J=3, \delta=2$, and $\omega= 0.0001$.}
	\label{threeparticles_pumping}
\end{figure}
The type-(i,iii) states are the three-particle bound state and scattering state.
With the initial state given by the multiparticle MLWS of the highest multiparticle Bloch band with Chern number $-1$, Fig.~\ref{threeparticles_pumping}(a) shows the density distribution $\langle \hat{n}_j(t) \rangle$ as a function of time in one cycle of time-dependent adiabatic evolution, where the red dashed line shows its mean position as a function of time. 
The three-particle bound state is unidirectionally shifted by $2.9759$ sites during a pumping cycle, consistent with the corresponding Chern number.
For the three-particle scattering states, The behaviors are quite different whether modulation is turned on or off. 
Fig.~\ref{threeparticles_pumping}(b) corresponds to the initial state given by the eigenstate of projected position operator of the highest scattering-state band under periodic boundary condition, and it is noticed that the state is extended.
Fig.~\ref{threeparticles_pumping}(c) shows a time-independent evolution for the same initial state in (b).
The dynamics is governed by $|\psi(t)\rangle=\mathcal T \exp[-i\int \hat H(t)dt]|\psi(0)\rangle$ for (a) and (b), and $|\psi(t)\rangle=\exp[-i\int \hat H(0)t]|\psi(0)\rangle$ for (c).
The parameters are $M=12, U_0=30, \delta=2, J=3, \omega= 0.0001$.
We find that particles in scattering state also have opportunity to interact, so the modulation of interaction can take effect on the three-particle scattering states. 

\section{Higher order correlation of the BIC and quasi-BIC}
We have shown the density distribution $\langle \hat{n}_j \rangle$ of two BICs under open boundary condition at $\phi=0$ with eigenvalues $\epsilon=321.9826$ and $\epsilon=320.8269$ and quasi-BICs assisted with the type-(ii) cluster bands at $t=0$ with the highest energy and middle energy under periodic boundary condition in the main text.
\begin{figure}[!h]
	\centering
	\includegraphics[scale=0.4]{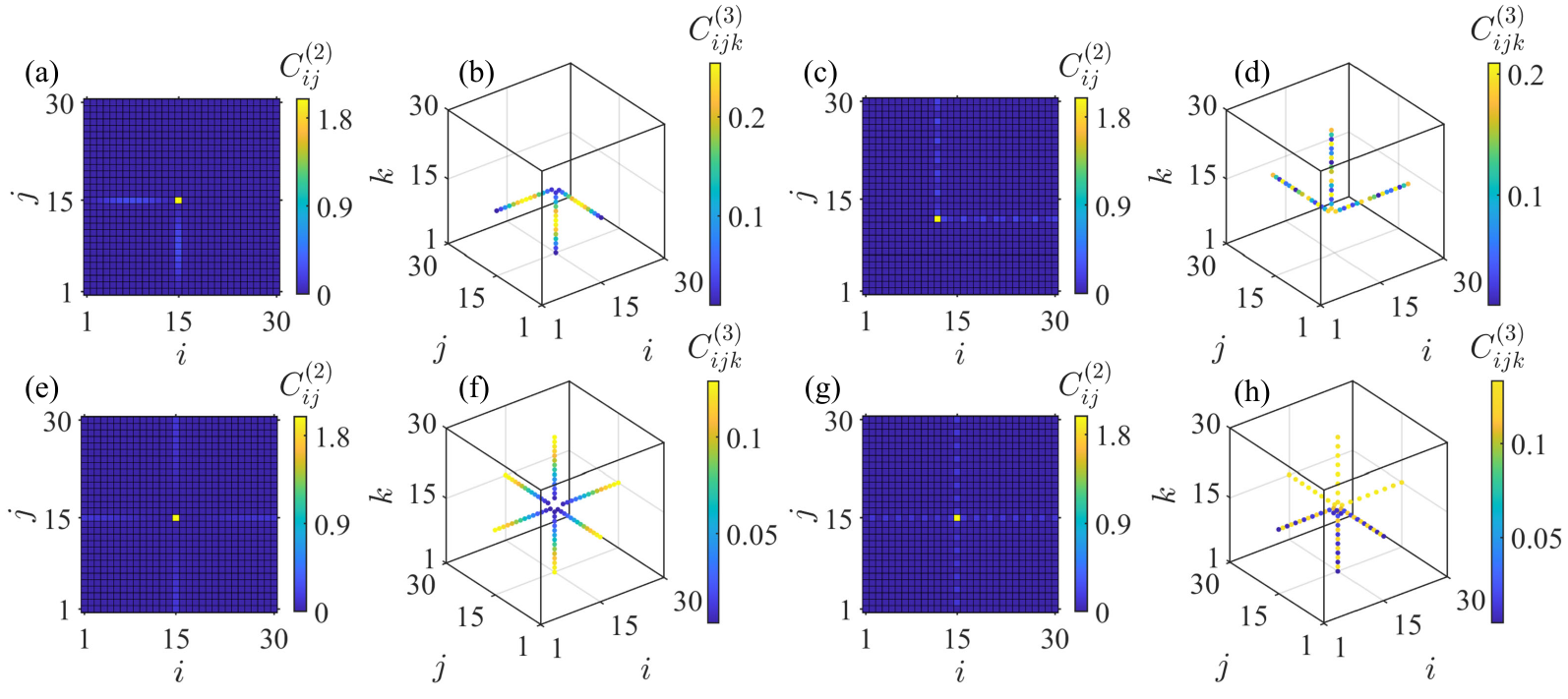}
	\caption{(a,c) the second-correlation and (b,d) the third correlation of two BICs at $\phi=0$ under open boundery condition with eigenvalues $\epsilon=321.9826$ and $\epsilon=320.8269$ respectively.
		(e,g) the second-correlation and (f,h) the third correlation of two quasi-BICs given by the the maximally localized Wannier state of the highest and middle type-(ii) bands at $t = 0$ under periodic boundery condition respectively.
		The parameters are $M=30, U_0=300, \delta=20, J=1$.}
	\label{highercorrelation}
\end{figure}
The density can be regarded as the first-order correlation of the states. 
Here, to get a full picture of the two states, we respectively calculate the second- and third-order correlation functions,
\begin{equation}
	\begin{aligned}
		& C_{ij}^{(2)}=\langle \psi | \hat{a}_{i}^{\dag}\hat{a}_{j}^{\dag}\hat{a}_j\hat{a}_i | \psi \rangle,\\
		& C_{ijk}^{(3)}=\langle \psi |\hat{a}_{i}^{\dag}\hat{a}_{j}^{\dag}\hat{a}_{k}^{\dag}\hat{a}_k\hat{a}_j\hat{a}_i |\psi \rangle.
	\end{aligned}
\end{equation}
Figs.~\ref{highercorrelation}(a,c) and (b,d) show the second- and third-correlation for the two BICs in the main text, and (e,g) and (f,h) for quasi-BICs in the main text, respectively. 
Parameters set as $M=30, U_0=300, \delta=20, J=1$. 
We can clearly see that the BICs are bound pair by the single-particle standing wave, while quasi-BICs are bound pair in the standing wave of the other particle.

\section{Demonstration of multiparticle MLWS as eigenstate of projected position operator}
A multiparticle MLWS can be obtained by minimizing its spread functional~\cite{PhysRevA.95.063630S}. A spread functional of multiparticle Wannier states in the $\mathcal{M}$ cluster band is written as
\begin{equation}
	\Omega=\sum_{m\in\mathcal{M}}\langle \hat{x}^2 \rangle_m-\langle \hat{x} \rangle^2_m,
\end{equation}
with $\langle \hat{x}^2 \rangle_m=\langle W_m(0)|\hat{x}^2|W_m(0)\rangle$ and $\langle \hat{x} \rangle_m=\langle W_m(0)|\hat{x}|W_m(0)\rangle$. The spread functional can be decomposed to two terms conveniently, 
\begin{equation}
	\Omega=\Omega_I+\Omega_V,
\end{equation}
where
\begin{equation}
	\begin{aligned}
		\Omega_I=\sum_{m\in\mathcal{M},n\notin\mathcal{M},R}|\langle  W_n(R)|\hat{x}|W_m(0)\rangle|^2,\\
		\Omega_V=\sum_{m,n\in\mathcal{M}}\sum_{Rn\neq 0m}|\langle W_n(R)|\hat{x}|W_m(0)\rangle|^2.
	\end{aligned}
\end{equation}
Here, It can be proved that $\Omega_I$ is gauge invariant under the unitary transformation $|u_m(\kappa)\rangle\rightarrow \sum_{n\in \mathcal{M}}U_{mn}(\kappa)|u_n(\kappa)\rangle$. So, the purpose of minimizing the spread functional is equal to minimizing $\Omega_V$. 
In one dimensional systems, a convenient method is to use the projected position operator $\hat{P}\hat{x}\hat{P}$, where
\begin{equation}
	\hat{P}=\sum_{m\in\mathcal{M},R}|W_m(R)\rangle\langle W_m(R)|=\sum_{m\in\mathcal{M},\kappa}|u_m(\kappa)\rangle\langle u_m(\kappa)|
\end{equation}
is the projected operator for the cluster of targeted multiparticle Bloch bands.
Here, $|W_m(R)\rangle$ and $|u_m(\kappa)\rangle$ represent a multiparticle Wannier state and multiparticle Bloch state, respectively.
Briefly, It can be seen that $\Omega_V$ vanishes when choosing the multiparticle Wannier states $|W_m(0)\rangle$ to be the eigenstate of the projected position operator $\hat{P}\hat{x}\hat{P}$ with an eigenvalue $X_{0m}$, that is
\begin{equation}
	\begin{aligned}
		\langle W_n(R)|\hat{x}|W_m(0)\rangle&=\langle W_n(R)|\hat{P}\hat{x}\hat{P}|W_m(0)\rangle
		\\&=X_{0m}\delta_{R,0}\delta_{m,n},
	\end{aligned}
\end{equation}
where $m,n\in \mathcal{M}$. So, by this way one can obtain the multiparticle MLWS. 

\section{Derivation of effective Hamiltonian for the BIC}
In this section, we show how to derive effective Hamiltonian for the BICs under open boundary condition, where singular value decomposition (SVD) and some semi-analytical derivation are used~\cite{Zhang_2023S}.  
A three-particle BIC $|\psi \rangle$ with a basis of Fock states can be reshaped to  $|\psi \rangle=\psi_{i,j,k}|\psi_{i,j,k}\rangle$, with the index $i,j,k\in [1,M]$ for the three particles. 
Due to the symmetry of the bosonic particles, any swap of $i,j,k$ does not change the amplitude $\psi_{i,j,k}$.
So, we have a tensor whose elements are $\psi_{i,j,k}$, which can be reshaped from a $M\times M\times M$ tensor to a $M\times M^2$ matrix.
After that, we denote the elements of the matrix as $\tilde{\psi}_{i,r}$, where $r=(j-1)\times M+k$. 
Performing SVD, there are two largest singular values that play the dominant role. 
Then, $\tilde{\psi}_{i,r}$ can be represented by two terms 
\begin{equation}
	\tilde{\psi}_{i,r}\approx D_{11}S_{i1}W_{1r}+D_{22}S_{i2}W_{2r}.
\end{equation}
$D_{11}$ and $D_{22}$ are the two largest  singular values, $S_{i1},S_{i2}$  represent a single-particle state, and $W_{1r},W_{2r}$ represent two-particle state. 
We denote localized single-particle state denoted as $S^{(l)}$, and the other extended state as $S^{(f)}$. 
For the two-particle state, we denote the two-particle localized state as $W^{(l,l)}$, and the state of one localized particle and one extended particle as $W^{(l,f)}$. 
Then, $\tilde{\psi}_{i,r}$ can be represented as
\begin{equation}
	\tilde{\psi}_{i,r}\approx D_{11}S^{(l)}_iW^{(l,f)}_r+D_{22}S^{(f)}_iW^{(l,l)}_r.
\end{equation}
We next reshape and perform SVD on $W^{(l,f)}$, and obtain a localized single-particle state $\mu^{(l)}$ and an extended state $\mu^{(f)}$,
\begin{equation}
	W^{(l,f)}_{j,k}\approx \mu^{(l)}_j\mu^{(f)}_k+\mu^{(l)}_k\mu^{(f)}_j.
\end{equation}
We denote $S^{(f)}\approx \mu^{(f)}$  as $\varphi^{(f)}$,  the rest localized part as $\chi^{(l)}$, and consider the exchange symmetry of bosons, the amplitudes of the there-particle BIC take the form of
\begin{equation}
	\psi_{i,j,k}=\varphi^{(f)}_i\chi^{(l)}_{j,k}+\varphi^{(f)}_j\chi^{(l)}_{i,k}+\varphi^{(f)}_k\chi^{(l)}_{i,j}.
\end{equation}
Motivated by this form, the three-particle Hamiltonian can be represented by 
\begin{equation}
	\hat{H}=\hat{H}^1\otimes I^2 \otimes I^3+I^1 \otimes \hat{H}^2 \otimes I^3+I^1\otimes I^2 \otimes H^3 +U, 
\end{equation}
where $\hat{H}$ with the superscript $1,2,3$ is linear hopping term, which works on the subspace of three individual particles, respectively. 
$I$ is the identity operator, and $U$ is the interacting term of $\hat{H}$. 
Submitting this form to stationary Schr\"{o}dinger equation, in the form of matrix operation, one can obtain 
\begin{equation}
	\sum_{i_1,i_2,i_3}(\hat{H}^1\otimes I^2 \otimes I^3+I^1 \otimes \hat{H}^2 \otimes I^3+I^1\otimes I^2 \otimes H^3+U)_{i_1,i_2,i_3}^{j_1,j_2,j_3}\psi_{i_1,i_2,i_3}=E\psi_{j_1,j_2,j_3}.
\end{equation}
Here, $i_1,i_2,i_3$ and $j_1,j_2,j_3$ are the matrix indices of the Hamiltonian for the first, second and third particles. 
The elements of the matrix of $U$ are given by
\begin{equation}
	U_{i_1,i_2,i_3}^{j_1,j_2,j_3}=\delta_{i_1,j_1}\delta_{i_2,j_2}\delta_{i_3,j_3}(U_{i_1}\delta_{i_1,i_2}+U_{i_3}\delta_{i_1,i_3}+U_{i_2}\delta_{i_2,i_3}),
\end{equation}
where $U_{i}=U_0+\delta\cos(2\pi\beta i+\phi)$.
Numerically, we find the single-particle extended state $\varphi^{(f)}$ is close to the eigenstate of $\hat{H}^1,\hat{H}^2,\hat{H}^3$, that is 
\begin{equation}
	\hat{H}^{1(2,3)}\varphi\approx\epsilon_n\varphi,
\end{equation}
where $\epsilon_n$ is corresponding eigenvalue. 
Hereafter, we omit the superscript of $\varphi$ and $\chi$ for brevity. 
Then, we can obtain
\begin{equation}
	\begin{aligned}
		&\epsilon_n\varphi_{j_1}\chi_{j_2,j_3}+\hat{H}^1_{i_1,j_1}\varphi_{j_2}\chi_{i_1,j_3}+\hat{H}^1_{i_1,j_1}\varphi_{j_3}\chi_{i_1,j_2}+\\
		&\hat{H}_{i_2,j_2}^2\varphi_{j_1}\chi_{i_2,j_3}+\epsilon_n\varphi_{j_2}\chi_{j_1,j_3}+\hat{H}_{i_2,j_2}^2\varphi_{j_3}\chi_{j_1,i_2}+\\
		&\hat{H}_{i_3,j_3}^3\varphi_{j_1}\chi_{j_2,i_3}+\hat{H}_{i_3,j_3}^3\varphi_{j_2}\chi_{j_1,i_3}+\epsilon_n\varphi_{j_3}\chi_{j_1,j_2}+\\
		&(\varphi_{j_1}\chi_{j_2,j_3}+\varphi_{j_2}\chi_{j_1,j_3}+\varphi_{j_3}\chi_{j_1,j_2})(U_{j_1}\delta_{j_1,j_2}+U_{j_3}\delta_{j_1,j_3}+U_{j_2}\delta_{j_2,j_3})\\
		&=E(\varphi_{j_1}\chi_{j_2,j_3}+\varphi_{j_2}\chi_{j_1,j_3}+\varphi_{j_3}\chi_{j_1,j_2}),
	\end{aligned}
\end{equation}
The same subscript of $i_1, i_2, i_3$ with $\hat H^{1,2,3}$ needs to be summed over this index. 
Next, multiplying the equation by $\varphi_{j_1}^{*}$ and summing over $j_1$, we have
\begin{equation}   \epsilon_n\chi_{j_2,j_3}+\hat{H}^2_{i_2,j_2}\chi_{i_2,j_3}+\hat{H}^3_{i_3,j_3}\chi_{j_2,i_3}+(2U_{j_2}|\varphi_{j_2}|^2\chi_{j_2,j_3}+2U_{j_3}|\varphi_{j_3}|^2\chi_{j_2,j_3}+U_{j_2}\chi_{j_2,j_3}\delta_{j_2,j_3})=E\chi_{j_2,j_3}.
\end{equation}
In the above derivation, we assume that $\varphi_{j_1}\chi_{j_2,j_3}=0$ if $j_1=j_2$, or $j_1=j_3$, or $j_1=j_2=j_3$. This is because $\chi$ should have little overlap to $\varphi$ for the BICs, and the energies of type-(ii) states have a large gap to those of type-(iii) states.
From the above equation, we can extract the effective Hamiltonian for $\chi$, 
\begin{equation}
	\hat{H}_{\rm eff}=J\sum_{j}(\hat{a}_{j+1}^{\dag}\hat{a}_j+h.c.)+\sum_{j}2U_j{|\varphi_j^{(f)}|}^2\hat{n}_j+\frac{1}{2}\sum_{j}U_j\hat{a}_{j}^{\dag}\hat{a}_{j}^{\dag}\hat{a}_j\hat{a}_j.
\end{equation}
Here, $U_j=U_0+\delta$cos$(2\pi\beta j+\phi)$, and $2U_j{|\varphi_j^{(f)}|}^2$ is the effective modulated on-site potential.

\section{Topological pumping of quasi-BIC in subspace}
\begin{figure}[!h]
	\centering
	\includegraphics[scale=0.3]{Subspace.pdf}
	\caption{Comparison of energy spectrum (a,b) and dynamics (c,d) between the total Hilbert space and the reduced Hilbert space expanded by type-(ii) Fock states. 
		The dashed line denotes the c.m. position of the bound pair. 
		The parameters are $M=12,U_0=300,J=1,\delta=20$ and $\omega= 0.0005$.}
	\label{Subspace}
\end{figure}

As the system size and particle number increase, the numerical simulation of dynamics costs enormous computational resources.   
For strong interaction leading to small correlation between the different types of states, we can simulate the dynamics in a subspace of a certain type which the initial state belongs.
The subspace of type-(ii) states is expanded by the basis $\mathcal V=\{|0,....,n_i=2,...,n_j=1,...\rangle \}$, where two bosons in the same site are bounded and the other boson at a different site.   
The dimension of this subspace is $M(M-1)$ for a $M$-sites lattice system, which is much less than $(M+2)(M+1)M/6$ of the total Hilbert space.
Figs.~\ref{Subspace}(a) and (b) show the energy spectrum of type-(ii) states as a function of time obtained in the full Hilbert space and type-(ii) subspace, respectively. 
Figs.~\ref{Subspace}(c) and (d) respectively show the density distribution $\langle \hat{n}_j \rangle$  as a function of time during a pumping cycle and in the total Hilbert space and type-(ii) subspace, where the initial state is the quasi-BIC that fills the highest type-(ii) state band.
The parameters are set $M=12, U_0=300, J=1, \delta=20$, and $\omega= 0.0005$. 
Both the energy spectrum and the dynamics are almost the same, indicating validation of the projected subspace method.
This method can also apply to the type-(i,iii) subspaces, and it is obviously beneficial to the numerical calculation of many-body systems. 

\section{Robustness against disorder for topological pumping of the quasi-BIC}
\begin{figure}[!h]
	\centering
	\includegraphics[scale=0.54]{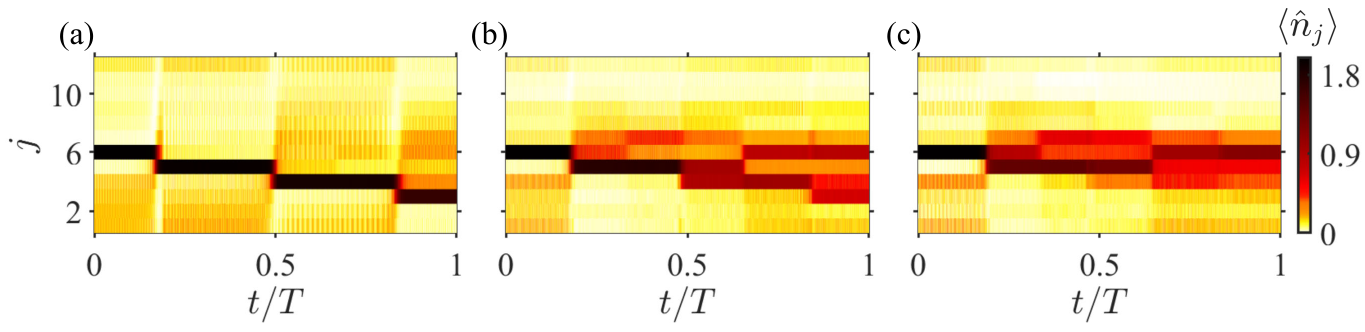}
	\caption{Density distribution as a function of time in a time-dependent adiabatic evolution with different disordered strengths $F=1,3,5$ for (a,b,c), respectively. The parameters are $M=12,U_0=300, \delta=20, J=1, \omega= 0.0005$, and $F=1$ for (a), $F=3$ for (b), $F=5$ for (c).}
	\label{disorder}
\end{figure}

In this section, we show that topological pumping of the quasi-BIC can be robust to disorder to some extent. 
We add an extra disordered onsite energies to the Hamiltonian, 
\begin{equation}
	\hat{H}=J\sum_{j}(\hat{a}_{j+1}^{\dag}\hat{a}_j+h.c.)+\frac{1}{2}\sum_{j}U_j(\phi)\hat{n}_j(\hat{n}_j-1)+\sum_{j}FV_j\hat{n}_j,
\end{equation}
where $V_j$ are random numbers in the range of $(0,1)$, and $F$ is the strength of disorder.
Without loss of generality, we take a group of random values of $V_j$ in numerical calculation,
\begin{equation}
	\{V_j\}=\{0.8308,0.5853,0.5497,0.9172,0.2858,0.7572,0.7537,0.3804,0.5678,0.0759,0.0540,0.5308\}.
\end{equation}
Setting the quasi-BIC filling the highest type-(ii) band without disorder as initial state, the time-dependent adiabatic evolution governed by $|\psi(t)\rangle=\mathcal T \exp[-i\int \hat H(t)dt]|\psi(0)\rangle$ is shown in Fig.~\ref{disorder}, where $F=1$ in (a), $F=3$ in (b) and $F=5$ in (c). 
The other parameters are set as $M=12,U_0=300, \delta=20, J=1, \omega= 0.0005$, and periodic boundary condition is adopted.
As can be seen, although it is a common trend that increasing disordered strength will break the topological pumping, with disorder strength to some extent such as $F=1$, the topological pumping is robust and persists well. 

\nocite{*}

\providecommand{\noopsort}[1]{}\providecommand{\singleletter}[1]{#1}%

\end{document}